\documentclass[aps, prd,  superscriptaddress, showpacs, longbibliography, floatfix, nofootinbib,10pt]{revtex4-2}

\usepackage{bm, amsmath, amssymb, relsize, amsfonts, mathrsfs, multirow, braket, siunitx, color, booktabs, arydshln, tabularx}
\usepackage{graphicx}
\graphicspath{{./figures/}}

\DeclareSIUnit \parsec {pc}

\DeclareUnicodeCharacter{202F}{\,} 

\usepackage[dvipsnames]{xcolor}
\usepackage[unicode]{hyperref}
\hypersetup{
  colorlinks=true,
  citecolor=RoyalBlue,
  linkcolor=blue,
  urlcolor=blue
}

\usepackage{array}
\usepackage{cancel}
\usepackage{orcidlink}
\usepackage{graphicx}
\usepackage{dcolumn}
\usepackage{bm}
\usepackage{paralist}
\usepackage{epsfig}
\usepackage{placeins}
\usepackage{mathrsfs}
\usepackage{comment}
\usepackage{color, colortbl}
\usepackage[inline]{enumitem}
\usepackage{soul}
\usepackage{adjustbox}
\usepackage{amssymb}
\usepackage{caption}
\usepackage{subcaption}
\usepackage{float}
\usepackage{cancel}
\usepackage{verbatim}
\usepackage{amsmath}
\usepackage[toc,page]{appendix}
\usepackage{graphicx}
\DeclareUnicodeCharacter{2212}{\ensuremath{-}}

\DeclareMathAlphabet{\mathpzc}{OT1}{pzc}{m}{it}
	
\definecolor{LightCyan}{rgb}{0.88,1,1}
\definecolor{lightgray}{gray}{0.9}

\def \IITGn     {Department of Physics, Indian Institute of Technology Gandhinagar, Gujarat 382055, India.\vspace*{4pt}}
\def \HRI     {Regional Centre for Accelerator-based Particle Physics, Harish-Chandra Research Institute, HBNI, Chhatnag Road, Jhunsi, Prayagraj (Allahabad) 211019, India\vspace*{4pt}}
\def \UoS     {School of Physics \& Astronomy, University of Southampton, Highfield, Southampton SO17 1BJ, UK.\vspace*{4pt}}
\def \UU     {Department of Physics \& Astronomy, Uppsala University, Box 516, 75120 Uppsala, Sweden.\vspace*{4pt}}

\begin{document}

\title{Explaining 650 GeV and 95 GeV Anomalies
in the 2-Higgs Doublet Model Type-I}

\author{\textsc{Akshat Khanna}\orcidlink{0000-0002-2322-5929}\vspace*{7pt}}
\email{khanna$_$akshat@iitgn.ac.in}
\affiliation{\IITGn}

\author{\textsc{Stefano Moretti}\orcidlink{0000-0002-8601-7246}\vspace*{7pt}}
\email{s.moretti@soton.ac.uk; stefano.moretti@physics.uu.se}
\affiliation{\UoS}
\affiliation{\UU}

\author{\textsc{Agnivo Sarkar}\orcidlink{0000-0001-9596-1936}}
\email{agnivosarkar@hri.res.in}
\affiliation{\HRI}


\begin{abstract}
We propose  an interpretation of a rather significant 650 GeV excess emerged at the Large Hadron Collider (LHC) from  CMS Collaboration data in the $\gamma\gamma b\bar b$ final state, accompanied by further clusters at 125(90-100) GeV in the $\gamma\gamma(b\bar b)$ system, within the 2-Higgs Doublet Model Type-I (2HDM-I) in presence of a softly broken $\mathcal{Z}_{2}$ symmetry.
The underlying process that we probe is $gg$-initiated production of a CP-odd (or pseudoscalar) Higgs boson $A$, with mass around 650 GeV, decaying into the Standard Model (SM)-like Higgs state $H$ (decaying into $\gamma\gamma$) and a $Z$ boson (decaying into $b\bar b$). We configure this theoretical framework so as to also have in the spectrum a light CP-even (or scalar) Higgs state $h$ with mass around 95 GeV, which is included for the purpose of simultaneously explaining additional data anomalies seen in the $b\bar b$, $\gamma\gamma$ and $\tau^+\tau^-$ final states while searching for light Higgs states at the Large Electron-Positron (LEP) collider (the first one) and LHC (the last two).  By accounting for both experimental and theoretical constraints, our results show that the 2HDM-I can explain all aforementioned anomalies at a significance level of $2.5 \sigma$.
\end{abstract}
\maketitle

\section{Introduction}
\label{sec:intro}
The detection at the Large Hadron Collider (LHC) of a 125 GeV Higgs boson in 2012~\cite{ATLAS:2012yve,CMS:2012qbp} has clearly verified that Electro-Weak  Symmetry Breaking (EWSB) is governed by the Higgs mechanism. This key event seems to have provided us with the only  missing element of the Standard Model (SM), as -- after many years of studying this object -- we have verified that its properties are very compatible with the corresponding predictions of such a theoretical framework. However, the SM, while  providing us with a comprehensive framework for particle physics, including both matter and interactions, 
it is also well-known to have several flaws (see Ref.~\cite{Khalil:2022toi} for a review). On the experimental side, it fails  in explaining phenomena like Dark Matter (DM), neutrino masses and the matter-antimatter asymmetry in the Universe. On the theoretical side, it suffers from the so-called hierarchy problem, i.e.,
its inability to reconcile the EW and Planck scales without excessive fine tuning of fundamental parameters. These shortcomings therefore motivate the exploration of possible extensions of the SM.

A possibility is to consider 2-Higgs-Doublet Models (2HDMs)~\cite{Gunion:1992hs,Branco:2011iw}, which predict the existence of additional (pseudo)scalar Higgs states  beyond the SM one. The search for these additional states, which could play a crucial role in addressing the above open questions, especially when 2HDMs are embedded in fundamental theories of the EW scale (like, e.g., Supersymmetry and Compositeness), has been ongoing for several years now.

In fact, the first evidence of the possible existence of one such additional particles emerged already at the Large Electron Positron (LEP) collider. Notably, the four experimental collaborations therein, ALEPH, DELPHI, L3 and OPAL (ADLO), have observed a moderate excess in the $e^+e^-\to Z^*\to Zb\bar{b}$ channel in the mass range of 95–100 GeV~\cite{ALEPH:2006tnd}, thus pointing to a possible CP-even Higgs state produced in association with the $Z$ (mostly decaying hadronically) and mediating the $b\bar b$ decay. More recently, several independent analyses at the LHC have reported potential signals of a light Higgs-like state near 95 GeV. Initially, 
CMS discussed an excess of approximately 2$\sigma$ in the di-photon (i.e., $\gamma\gamma$) channel near 97 GeV during Run 1~\cite{CMS:2015ocq}, which was later confirmed by Run 2 data, which provided a local significance of 2.9$\sigma$ at $m_{\gamma\gamma}\approx 95$ GeV \cite{CMS:2018cyk,CMS:2024yhz}. Furthermore, an independent search by ATLAS in the di-photon channel observed a similar excess with a local significance of 1.7$\sigma$~\cite{ATLAS:2024bjr,ATLAS:2023jzc}. Finally, CMS additionally observed a 2.6$\sigma$ excess, also near 95 GeV, in the di-tau
(i.e., $\tau^+\tau^-$) channel~\cite{CMS:2022goy}.

Even more intriguingly, a very recent CMS search in the $\gamma\gamma b\bar{b}$ channel has reported quite a significant excess at approximately 650 GeV, with a local significance of 3.8$\sigma$~\cite{CMS:2023boe}. This excess has been interpreted as evidence of a heavy resonance decaying into a SM-like Higgs boson (of 125 GeV) and an additional state, potentially (but not necessarily) a Higgs boson with a mass of about 95 GeV. The measured production cross section  of the possible 650 GeV resonance  times the relevant Branching Ratios (${\rm BR}$s), i.e., of the relevant decays into $\gamma\gamma$ and $b\bar b$, 
is approximately 0.35 fb.

The simultaneous observation of all such excesses  has gathered widespread  interest in theoretical models that can possibly explain these, separately or together. Here is an exhaustive list of references: \cite{Cao:2016uwt,Heinemeyer:2021msz,Biekotter:2021qbc,Biekotter:2019kde,Cao:2019ofo,Biekotter:2022abc,Iguro:2022dok,Li:2022etb,Cline:2019okt,Biekotter:2021ovi,Crivellin:2017upt,Cacciapaglia:2016tlr,Abdelalim:2020xfk,Biekotter:2022jyr,Biekotter:2023jld,Azevedo:2023zkg,Biekotter:2023oen,Cao:2024axg,Wang:2024bkg,Li:2023kbf,Dev:2023kzu,Borah:2023hqw,Cao:2023gkc,Aguilar-Saavedra:2023tql,Ashanujjaman:2023etj,Dutta:2023cig,Ellwanger:2024txc,Diaz:2024yfu,Ellwanger:2024vvs,Ayazi:2024fmn,Coloretti:2023wng,Bhattacharya:2023lmu,Ahriche:2023hho,Ahriche:2023wkj,Benbrik:2022azi,Benbrik:2022tlg,Ellwanger:2023zjc,Banik:2023ecr,Belyaev:2023xnv,Janot:2024lep,Gao:2024ljl,Benbrik:2024ptw,Li:2025tkm,Hmissou:2025uep,Gao:2024qag,Dutta:2025nmy,Abbas:2025ser,Xu:2025vmy,Arhrib:2025pxy,Coutinho:2024zyp,Abbas:2024jut,Baek:2024cco,Banik:2024ugs,Mondal:2024obd,Dong:2024ipo,Robens:2024wbw,BrahimAit-Ouazghour:2024img,Khanna:2024bah,Janot:2024ryq,YaserAyazi:2024hpj,Ogreid:2024qfw,Du:2025eop,Lian:2024smg,Benbrik:2025hol,Hammad:2025wst,Kundu:2024sip}. Among these extensions, as evident from sampling such a list, 
the 2HDM remains particularly popular, specifically, because of  its minimality (only one additional doublet with respect to the SM) and its ability to naturally accommodate additional (pseudo)scalar Higgs particles. 
In particular, it has been proven that in its Type-III incarnation, the 2HDM can simultaneously explain both the 95 and 650 GeV excesses \cite{Benbrik:2025hol}.

Trailing this last paper, here, we study instead the 
2HDM Type~I (2HDM-I) scenario, where a heavy CP-odd Higgs boson, $A$, with a mass near 650 GeV decays into a SM-like Higgs boson, $H$, and a $Z$ boson in the decay channel $A \to HZ$. The SM-like Higgs boson $H$ then decays into a di-photon pair whereas the $Z$ boson decays into a $b\bar{b}$ one. In fact, we also retain in the particle spectrum of the 2HDM-I the $h$ state fitting the aforementioned low mass excesses.  Notice that this interpretation is well motivated. In fact, although the CMS analysis~\cite{CMS:2023boe} originally models the excess as a decay $X_{650} \to h^{125}_{\rm SM} Y$, with $h^{125}_{\rm SM}$($Y$) being the SM-like Higgs boson(a new spin-0 state), at $\sqrt s =m_A \simeq 650$~GeV, where $m_Z^2/s \ll 1$, the   equivalence theorem ensures that the psudoscalar polarisation of the $Z$ boson behaves like the corresponding neutral Goldstone mode, indeed, a spin-0 state, with deviations 
(due to the transverse polarizations states  of the $Z$ boson) suppressed by ${\cal O}(m_Z^2/s) \lesssim 2\%$. In fact, as seen in  previous published literature \cite{Benbrik:2025hol, Hmissou:2025riw}, wherein such a solution
was invoked, a MC analysis of $Z\to b\bar b$ decays was performed and no significant difference was found in the signal sensitivity to the CMS cuts between the transverse ($\pm1$) and longitudinal ($0$) polarizations of the $Z$ boson. There is, however, a difference between the $Z$ interpretation and the one using a Higgs boson for the ${M_{b\bar b}}$ excess, i.e.,  the $Z$ boson is quite wide  ($\Gamma_Z/M_Z\approx 3\%$) compared to possible (much narrower) spin-0 explanations: this required a correction to the normalization of the event rates for our process, given that CMS starts sampling the $b\bar b$ invariant mass from 70 GeV, which we have accounted for in all our results.

Hence, also considering the limited mass resolution of $b\bar b$ pairs, of some 10 GeV or so, and the fact that none of the selection cuts used by CMS has a marked spin dependence, it is conceivable that the $Z$ boson of the SM could be  behind the 95 GeV component of the 650 GeV anomaly. 

The rest of this paper is organised as follows. In Sect.~\ref{sec:model}, we detail the theoretical framework of the CP-conserving 2HDM-I. The following section discusses both the theoretical and experimental constraints that must be enforced on the parameter space of our Beyond the SM (BSM) framework. Then, in Sect.~\ref{sec:excess}, 
we present the results of our numerical analysis. We finally conclude.
\section{The 2HDM-I}
\label{sec:model}
In this section we will briefly review the scalar sector of 2HDM-I. To avoid tree level Flavour Changing Neutral Current (FCNC) processes, one can assume an additional discrete symmetry $\mathcal{Z}_{2}$ in addition to the SM gauge symmetries, $SU(2)\times  U(1)_{Y}\times SU(3)_C$, under which the doublet fields transform as $\phi_{1} \to \phi_{1}$ and $\phi_{2} \to -\phi_{2}$. Alongside the (pseudo)scalar fields, the SM fermion fields are also charged under this $\mathcal{Z}_{2}$ symmetry and transform as $u^{i}_{R} \to -u^{i}_{R}$ (right handed up quark), $d^{i}_{R} \to -d^{i}_{R}$ (right handed down quark) and $\ell^{i}_{R} \to -\ell^{i}_{R}$ (right handed lepton), respectively. With these charge assignments, one can realise that the fermion fields  exclusively couple to the $\phi_{2}$ doublet. In Eq.(\ref{eq:scalarpot}) we present the Higgs potential of the model.

\begin{equation}
    \label{eq:scalarpot}
    \begin{split}
    V(\phi_1, \phi_2) & = m_{11}^2 \phi_1^\dagger \phi_1 + m_{22}^2 \phi_2^\dagger \phi_2 - m^{2}_{12}\left[\phi^{\dagger}_{1}\phi_{2} + \phi^{\dagger}_{2}\phi_{1}\right] + \frac{\lambda_1}{2}(\phi_1^\dagger \phi_1)^2 + \frac{\lambda_2}{2}(\phi_2^\dagger \phi_2)^2 \\
    & + \lambda_3(\phi_1^\dagger \phi_1)(\phi_2^\dagger \phi_2) + \lambda_4(\phi_1^\dagger \phi_2)(\phi_2^\dagger \phi_1) + \{   \frac{\lambda_5}{2}(\phi_1^\dagger \phi_2)^2 + h.c. \}.
    \end{split}
\end{equation}

Herein, one observes that the term proportional to $m^{2}_{12}$ introduces a soft breaking of the $\mathcal{Z}_{2}$ symmetry. This soft breaking plays an important role in enlarging the viable parameter space, allowing for parameter points that satisfy the theoretical consistency conditions. Both  doublet fields, $\phi_{1}$ and $\phi_{2}$, acquire Vacuum Expectation Values (VEVs) and break the underlying gauge symmetry down to the $U(1)_{\rm  EM}$ one of the Electro-Magnetic (EM) sector. After electroweak symmetry breaking, the scalar doublets are expanded around their vevs which leads to mass matrices in the charged, CP-odd, and CP-even scalar sectors. The charged and CP-odd mass matrices are diagonalized by the angle $\beta$, defined through $\tan\beta=v_2/v_1$, yielding the physical states $H^\pm$ and $A$. The CP-even mass matrix is diagonalized by an additional mixing angle $\alpha$, giving rise to the physical CP-even states $h$ and $H$. (with $m_h<m_H$). Hereafter, we identify $H$ as the SM-like Higgs boson and $h$ as a lighter state. Expressing the entries of these mass matrices in terms of the potential parameters $\lambda_i$, $m_{12}^2$, and the VEVs, and inverting these relations, one obtains the expressions of the quartic couplings $\lambda_i$ in terms of the physical scalar masses, mixing angles, and $\tan\beta$, as summarized in Eq.~(\ref{Eq:lambda}). The detailed derivations can be found in \cite{Branco:2011iw,Bhattacharyya:2015nca}.

\begin{equation}\label{Eq:lambda}
    \begin{split}
    \lambda_1 & = \frac{c_\alpha^2m_{H}^2 + s_\alpha^2m_{h}^2}{v^2c_{\beta}^2} - \frac{m_{12}^2s_{\beta}}{v^2c_{\beta}^3}, \\
    \lambda_2 & = \frac{c_\alpha^2m_{h}^2 + s_\alpha^2m_{H}^2}{v^2s_{\beta}^2} - \frac{m_{12}^2c_{\beta}}{v^2s_{\beta}^3}, \\
    \lambda_3 & = \frac{(m_{H}^2-m_{h}^2)s_{\alpha}c_{\alpha} - (\lambda_4+\lambda_5)v^2c_{\beta}s_{\beta}}{v^2c_{\beta}s_{\beta}} + \frac{m_{12}^2}{v^2c_{\beta}s_{\beta}}, \\
    \lambda_4 & = \frac{m_A^2 - 2m_{H^\pm}^2}{v^2} + \frac{m_{12}^2}{v_1v_2}, \\
    \lambda_5 & = -\frac{m_A^2}{v^2} + \frac{m_{12}^2}{v_1v_2}.
    \end{split}
\end{equation}
Comparing with Eq.(\ref{eq:scalarpot}), one can notice that the potential has seven independent parameters. Writing down the above potential in the mass basis,  instead, one can express all the $\lambda_i$'s in terms  of the physical Higgs masses $h, H, A$ and $H^\pm$, the angles $\alpha$ (regulating the mixing between the two neutral CP-even Higgs states) and $\beta$ (the ratio of the two VEVs) plus the softly symmetry breaking term $m_{12}$. We use the latter parameterisation for further analysis.

In Eq.~(\ref{eq:yukeq}) we write down the Yukawa part of the Lagrangian in the mass eigenstate basis:  

\begin{equation}
		\label{eq:yukeq}
  \begin{split}
      - \mathcal{L}_{\rm Yukawa} & = +\sum_{f = u,d,\ell} \left[ m_{f}f\bar{f} + \left(\frac{m_{f}}{v}\kappa^{f}_{h}\bar{f}fh + \frac{m_{f}}{v}\kappa^{f}_{H}\bar{f}fH - i\frac{m_{f}}{v}\kappa^{f}_{A}\bar{f}\gamma_{5}fA\right)\right] \\
             & ~ + \frac{\sqrt{2}}{v}\bar{u}\left(m_{u}V\kappa^{u}_{H^{+}}P_{L} + Vm_{d}\kappa^{d}_{H^{+}P_{R}}\right)dH^{+} + \frac{\sqrt{2}m_{\ell}\kappa^{\ell}_{H^{+}}}{v}\bar{\nu}_{L}\ell_{R}H^{+} + h.c.
  \end{split}	
\end{equation}
Here $m_{f}$ is the fermion mass, $V$ is the Cabibbo-Kobayashi-Maskawa (CKM) matrix and $P_{R/L} = \frac{1 +/- \gamma_{5}}{2}$ are the left- and right-handed projection operators. The explicit form of the scaling functions $\kappa_{i}$'s (also called coupling modifiers) are detailed in Table~\ref{tab:couplings}.

\begin{table}[!ht]
    \centering
    \begin{tabular}{l l}
    \toprule[1pt]
        $\kappa^{i}_{S}$ & \ \ \ \ \ \ \ Coefficient \\
        \midrule[1pt]
        $\kappa^{V}_{H}$ & \ \ \ \ \ \ \ $\cos(\beta - \alpha)$ \\
        $\kappa^{V}_{h}$ & \ \ \ \ \ \ \ $\sin(\beta - \alpha)$ \\
        $\kappa^{f}_{H}$ & \ \ \ \ \ \ \ $\frac{\sin \alpha}{\sin \beta}$ \\
        $\kappa^{f}_{h}$ & \ \ \ \ \ \ \ $\frac{\cos \alpha}{\sin \beta}$ \\
        $\kappa^{f}_{A}$ & \ \ \ \ \ \ \ $\cot\beta$ \\
        $\kappa^{u}_{H^{+}}$ & \ \ \ \ \ \ \ $\cot\beta$ \\
        $\kappa^{d/\ell}_{H^{+}}$ & \ \ \ \ \ \ \ $-\cot\beta$ \\
    \bottomrule[1pt]
    \bottomrule[1pt]
    \end{tabular}
    \caption{Explicit form of different coupling modifiers $\kappa^{i}_{S}$. Here $S$ denotes different (pseudo) scalar states in the 2HDM-I and $i$ can refer to SM gauge bosons and fermions \cite{Branco:2011iw}.}
    \label{tab:couplings}
\end{table}

\section{Constraints}
\label{sec:constraints}
In this section, we describe different theoretical and experimental constraints which are required to restrict the parameter space of the 2HDM-I. 

\label{sec:constraints}

\subsection{Theoretical Constraints}

\begin{itemize}
    \item \textbf{Vacuum Stability:} 
    To satisfy the vacuum stability one needs to check whether the potential is strictly positive for all possible large field directions. To achieve this, the $\lambda_{i}$ parameters must follow certain relationships such that the quartic terms in the potential must dominate for large field values. In the following we write down the conditions on the $\lambda_i$'s that ensure the stability criteria  \cite{Coleppa_2014}:
    \begin{equation*}
    \lambda_1  > 0, \,\,\,\,\, \lambda_2  > 0,  \,\,\,\,\, \lambda_3 + \sqrt{\lambda_1 \lambda_2}  > 0,  \,\,\,\,\, \lambda_3 + \lambda_4 - |\lambda_5| + \sqrt{\lambda_1 \lambda_2}  > 0.
     \end{equation*}
    
    \item \textbf{Unitarity:} The requirement of unitarity places important restrictions on the parameter space of the model, ensuring that scattering amplitudes remain well-behaved at high energies. At tree level, this translates into conditions on the eigenvalues of the full set of $2 \to 2$ (pseudo)scalar scattering processes involving Higgs and Goldstone bosons. For the 2HDM, these conditions have been worked out explicitly in Refs.~\cite{Ginzburg_2005,Bhattacharyya:2015nca}. The resulting unitarity bounds take the following form:
    \begin{equation*}
         |u_i| \leq 16\pi~(i=1,... 6),
    \end{equation*}
    where
     \begin{equation*}
    \begin{split}
    u_1 & = \frac{1}{2}(\lambda_1 + \lambda_2 \pm \sqrt{(\lambda_1 - \lambda_2)^2 + 4|\lambda_5|^2}), \\ 
    u_2 & = \frac{1}{2}(\lambda_1 + \lambda_2 \pm \sqrt{(\lambda_1 - \lambda_2)^2 + 4\lambda_4^2}), \\
    u_3 & = \frac{1}{2}(3(\lambda_1 + \lambda_2) \pm \sqrt{9(\lambda_1 - \lambda_2)^2 + 4(2\lambda_3+\lambda_4)^2}), \\
    u_4 & = \lambda_3 + 2\lambda_4 \pm 3|\lambda_5|, \\
    u_5 & = \lambda_3 \pm |\lambda_5|, \\
    u_6 & = \lambda_3 \pm \lambda_4. \\
    \end{split}
\end{equation*}
    \item \textbf{Perturbativity:} Finally, we also demand that the model remains within the perturbative regime. To maintain this condition,  one can impose further upper bounds on $\lambda_{i}$'s corresponding to the quartic terms of the potential: $\lambda_i \leq |8\pi|$ ($i=1,... 5$).
\end{itemize}

\subsection{Experimental Constraints}

\begin{itemize}
    \item \textbf{EW Precision Tests:} We evaluated the EW precision constraints by computing the $S, T$ and $U$ parameters using the SPheno package \cite{Porod_2003}, with the model files written in SARAH  \cite{Staub_2015}. These so-called `oblique parameters' provide stringent constraints on new physics masses and relevant couplings. Therefore any BSM scenario should conform to these precision data which were primarily collected by LEP, SLC and  Tevatron. In the present scenario, these constraints typically enforce an approximate degeneracy between the CP-odd neutral state and the charged ones. The numerical values, with correlation coefficients of $+0.92$ between $S$ and $T$ plus $-0.68$ ($-0.87$) between $S$ and $U$ ($T$ and $U$) are \cite{2018}
\begin{equation*}
    S = 0.04 \pm 0.11, \,\,\,\,\,\,\, T = 0.09 \pm 0.14, \,\,\,\,\,\,\, U  = -0.02 \pm 0.11.         
\end{equation*}

    \item \textbf{BSM Higgs Boson Exclusions:} We assessed the exclusion limits from direct searches for  BSM (pseudo)scalar states at  LEP, Tevatron and the LHC. These exclusion limits were evaluated at the $95 \%$ Confidence Level (C.L.) using the HiggsBounds-6 \footnote{The analysis is performed using the HiggsBounds dataset v1.6.}\cite{Bechtle_2020} module via the HiggsTools \cite{Bahl_2023} package. In our analysis, we have also demanded that our lighter Higgs boson must comply with the results of ~\cite{CMS:2024ulc}, where the Higgs particles are produced in association with a massive vector boson or a top anti-quark pair and further decays via leptonic modes.

    \item \textbf{SM-Like Higgs Boson Discovery:} We then examined the compatibility of our 125 GeV Higgs boson $H$ with the discovered SM-like Higgs boson using a goodness-of-fit test. Specifically, we calculated the $\chi^2$ value with HiggsSignals-3 \footnote{The analysis is performed using the HiggsSignals dataset v1.1.} \cite{Bechtle_2021} via HiggsTools, comparing the predicted signal strengths of our Higgs boson to those observed experimentally. We retained the parameter spaces that satisfies the conditions $\chi_{125}^2 < 166.93$, $\chi_{125}^2 < 189.42$ and $\chi_{125}^2 < 213.11$, corresponding to $68.27 \%$, $95 \%$ and $99.73 \%$ C.L., respectively, with $159$ degrees of freedom \cite{Benbrik:2024ptw, Belyaev:2023xnv}. The choice of these upper bounds corresponds to a global $\chi^2$ goodness-of-fit test ensuring that the predicted properties of the heavier CP-even Higgs state (with  mass $m_{H} = 125$ GeV) of the underlying 2HDM-I closely match those of the Higgs boson observed at the LHC, thereby keeping the model compatible with these experimental constraints too.

    \item \textbf{Flavour Physics:} We incorporated constraints from $B$-physics observables, which are sensitive to potential new physics contributions in loop mediated FCNC processes, specifically, we tested the most stringent bound on the $B\rightarrow X_s \gamma$ decay using Next-to-Leading Order (NLO) calculations in QCD as discussed in \cite{Borzumati_1998}:
    \begin{equation}
        {\rm BR}(B \rightarrow X_s \gamma) = \frac{\Gamma (B \rightarrow X_s \gamma)}{\Gamma_{\rm SL}}{\rm BR}_{\rm SL},
    \end{equation}
    where ${\rm BR}_{\rm SL}$ is the semi-leptonic ${\rm BR}$ and $\Gamma_{\rm SL}$ is the semi-leptonic decay width. These can impose a stringent bound on the charged Higgs boson mass as well as its coupling with SM fermions.
    %
%
%
    We took our input parameters from the most recent Particle Data Group (PDG) compilation \cite{Workman:2022ynf}, as follows:
    \begin{align*}
         \alpha_s(M_Z) & = 0.1179 \pm 0.0010 , \ \ \ \ \ \ \ \ \ \ \ \ \ \ m_t = 172.76 \pm 0.3, \\
         \frac{m_b}{m_c} & = 4.58 \pm 0.01, \ \ \ \ \ \ \ \ \ \ \ \ \ \ \alpha  = \frac{1}{137.036}, \\
        {\rm BR}_{\rm SL} & = 0.1049 \pm 0.0046, \ \ \ \  |\frac{V_{ts}^*V_{tb}}{V_{cb}}|^2 = 0.95 \pm 0.02, \\
         m_b(\overline{\rm MS}) & = 4.18 \pm 0.03, \ \ \ \ \ \ \ \ \ \ \ \ \ \ m_c ({\overline{\rm MS}}) = 1.27 \pm 0.02, \\
        m_Z & = 91.1876 \pm 0.0021, \ \ \ \ \ \ \ \ \  m_W = {80.36} \pm 0.012.
    \end{align*}
    The following restriction has then been imposed, which represents the $3 \sigma$ experimental limit: 
    \begin{equation*}
         2.87 \times 10^{-4} < {\rm BR}(B \rightarrow X_s \gamma) < 3.77 \times 10^{-4}.
    \end{equation*}
    Other $B$-physics observables, like ${\rm BR}(B^+ \rightarrow \tau^+ \nu_\tau)$, ${\rm BR}(D_s \rightarrow \tau \nu_\tau)$, ${\rm BR}(B_s \rightarrow \mu^+ \mu^-)$ and  ${\rm BR}(B^0 \rightarrow \mu^+ \mu^-)$ have been taken care of by using the FlavorKit tool \cite{Porod_2014} provided by the SPheno package  \cite{Porod_2003}. Our calculated $b \rightarrow s \gamma$ results were also found to be consistent with the FlavorKit tool.
\end{itemize}

\section{Analysing the Excesses}
\label{sec:excess}
In this section, we illustrate how to explain the excesses observed at $650$ GeV and $95$ GeV in the $\gamma \gamma b\bar{b}$ as well as $\gamma\gamma$, $\tau^+\tau^-$ and $b\bar b$ channels, respectively. In particular, as mentioned, the first (combined) excess can be interpreted in the 2HDM-I as a heavy CP-odd scalar decaying into a SM-like Higgs boson $H$ accompanied by a $Z$ boson, i.e., $gg\to A\to H Z\to \gamma\gamma b\bar b $. The CMS collaboration reports the best-fit signal yield of
\begin{equation}    \sigma^{\rm CMS}_{\gamma \gamma b \bar{b}} = 0.35^{+0.17}_{-0.13} \; \text{fb}.
\end{equation}
As intimated, the aforementioned additional state $Y$ can be interpreted as either  a lighter Higgs boson or a $Z$ boson. In this work, we treat $Y$ as the $Z$ boson, which allows the observed excess around 95 GeV to be uniquely  attributed to the light CP-even state, $h$, while $X$ corresponds to the CP-odd Higgs, $A$. To characterise the low-mass excess at 95 GeV, we define the signal strengths in the $\tau^+\tau^-$, $\gamma\gamma$ and $b\bar{b}$ final states as
\begin{equation}
    \begin{split}
        \mu_{\tau^+ \tau^-} & = \frac{\sigma_{\rm 2HDM}(gg\rightarrow H)}{\sigma_{\rm SM}(gg\rightarrow h^{95}_{\rm SM})} \times \frac{{\rm BR}_{\rm 2HDM}(H \rightarrow \tau^+ \tau^-)}{{\rm BR}_{\rm SM}(h^{95}_{\rm SM} \rightarrow \tau^+ \tau^-)}, \\
        \mu_{\gamma \gamma} & = \frac{\sigma_{\rm 2HDM}(gg\rightarrow H)}{\sigma_{\rm SM}(gg\rightarrow h^{95}_{\rm SM})} \times \frac{{\rm BR}_{\rm 2HDM}(H \rightarrow \gamma \gamma)}{{\rm BR}_{\rm SM}(h^{95}_{\rm SM} \rightarrow \gamma \gamma)}, \\
        \mu_{b \bar{b}} & = \frac{\sigma_{\rm 2HDM}(e^+e^-\rightarrow Z H)}{\sigma_{\rm SM}(e^+e^-\rightarrow Z h^{95}_{\rm SM})} \times \frac{{\rm BR}_{\rm 2HDM}(H \rightarrow b \bar{b})}{{\rm BR}_{\rm SM}(h^{95}_{\rm SM} \rightarrow b \bar{b})}.
    \end{split}
    \label{Eq:sigstrength}
\end{equation}
Here, $h^{95}_{\rm SM}$ corresponds to a SM-like Higgs Boson with a mass of 95 GeV while $h$ is a 2HDM-I Higgs state with the same mass. The experimental measurements for these three signal strengths around $95$ GeV are expressed as 
\begin{equation}
    \begin{split}
        \mu_{\gamma \gamma}^{\rm exp} & = \mu_{\gamma \gamma}^{\rm ATLAS+CMS} = 0.24^{+0.09}_{-0.08}~\text{\cite{CMS-PAS-HIG-20-002, CMS:2018cyk, ATLAS-CONF-2023-035}},  \\
        \mu_{\tau^+ \tau^-}^{\rm exp} & = 1.2 \pm 0.5~\text{\cite{CMS:2022goy}}, \\
        \mu_{b \bar{b}}^{\rm exp} & = 0.117 \pm 0.057~\text{\cite{LEPWorkingGroupforHiggsbosonsearches:2003ing, Cao_2017}}.
    \end{split}
    \label{eq:signal_excess}
\end{equation}
The contribution to the $\chi^2$ for each of the channels is calculated using the formula 
\begin{equation}
    \chi^2_{\tau^+ \tau^-, \gamma \gamma, b \bar{b}} = \frac{(\mu_{\tau^+ \tau^-, \gamma \gamma, b \bar{b}} - \mu_{\tau^+ \tau^-, \gamma \gamma, b \bar{b}}^{\rm exp})^2}{(\Delta \mu_{\tau^+ \tau^-, \gamma \gamma, b \bar{b}}^{\rm exp})^2}.
\end{equation}
Hence, the resulting $\chi^2$, which we will use to determine if the excesses can be achieved by any (theoretically and experimentally) allowed region of the parameter space of the 2HDM-I, or otherwise, is the following:
\begin{equation}
    \label{eq:chisum}
    \chi^2_{\tau^+ \tau^- + \gamma \gamma +  b \bar{b}} =  \chi^2_{\ \tau^+ \tau^-} + \chi^2_{\gamma \gamma} + \chi^2_{b \bar{b}}.
\end{equation}
As a reference, a 2HDM-I parameter point is considered consistent with the data at the $1 \sigma$ level if $\chi^2_{\tau^+ \tau^- + \gamma \gamma +   b \bar{b}}<3.52$.

\section{Explaining the anomalies}
\label{sec:anomaly}
In this section, we present our numerical analysis to investigate whether the 2HDM-I can simultaneously account for the excesses observed at $650$ GeV and $95$ GeV. The model file was generated using the \texttt{SARAH} package \cite{Staub_2015}, while the spectrum generator \texttt{SPheno} \cite{Porod_2003} was employed to compute the cross sections and BRs. Monte Carlo (MC) samples were generated within the parameter ranges specified in Table~\ref{tab:scanrange} and subsequently tested against the algebraic inequalities corresponding to vacuum stability, perturbative unitarity, and perturbativity using a python based implementation and experimental constraints using the \texttt{HiggsTools} package. The generated points were further tested against the most recent CMS upper limits on the cross section for the process $pp\to Y(\to\tau^+\tau^-)H(\to\gamma\gamma)$ \cite{CMS:2025tqi} and these were found not constraining the relevant regions of 2HDM-I parameter space. In fact, bearing in mind \cite{CMS:2025qit} alongside \cite{CMS:2025tqi}, the following production and decay patterns were also tested: $pp\to H(\to b\bar b)Y(\to\gamma\gamma)$ and $pp\to H(\to\tau^+\tau^-)Y(\to\gamma\gamma)$. However, since in our case $Y\equiv Z,$\footnote{And the 2HDM-I used here is CP-conserving, so that the decay $A\to Hh$ is not possible.} the (non-resonant) transition $Z^*\to\gamma\gamma $  is highly suppressed in view of the Landau-Yang theorem
\cite{Landau:1948kw,Yang:1950rg} (see also \cite{Moretti:2014rka}),  the corresponding experimental limits are inconsequential for our theoretical scenario.


We thus work within the inverted mass hierarchy scenario of the 2HDM-I, where $H$ corresponds to the SM-like Higgs boson (with a mass of exactly 125 GeV) and the lighter CP-even state $h$ lies in the range $93$–$97$ GeV. This lighter state is responsible for the excesses observed near $95$ GeV in the $\gamma\gamma$, $\tau^+\tau^-$ and $b\bar{b}$ channels. At the same time, the CP-odd scalar $A$ is taken in the range $625$–$665$ GeV, decaying predominantly through $A \rightarrow HZ$, followed by $H \rightarrow \gamma \gamma$ (as computed by {\tt SPheno}) and $Z \rightarrow b \bar{b}$ (${\rm BR}=0.15$). The BR of the diphoton channel lies in the range $0.0024$–$0.0035$, as we relax the alignment limit and vary $\sin(\beta-\alpha)$ during parameter sampling. Hence, this decay chain provides a natural explanation for the anomaly reported around $650$ GeV. 

\begin{figure}[h!]
    \centering
    \includegraphics[scale=0.45]{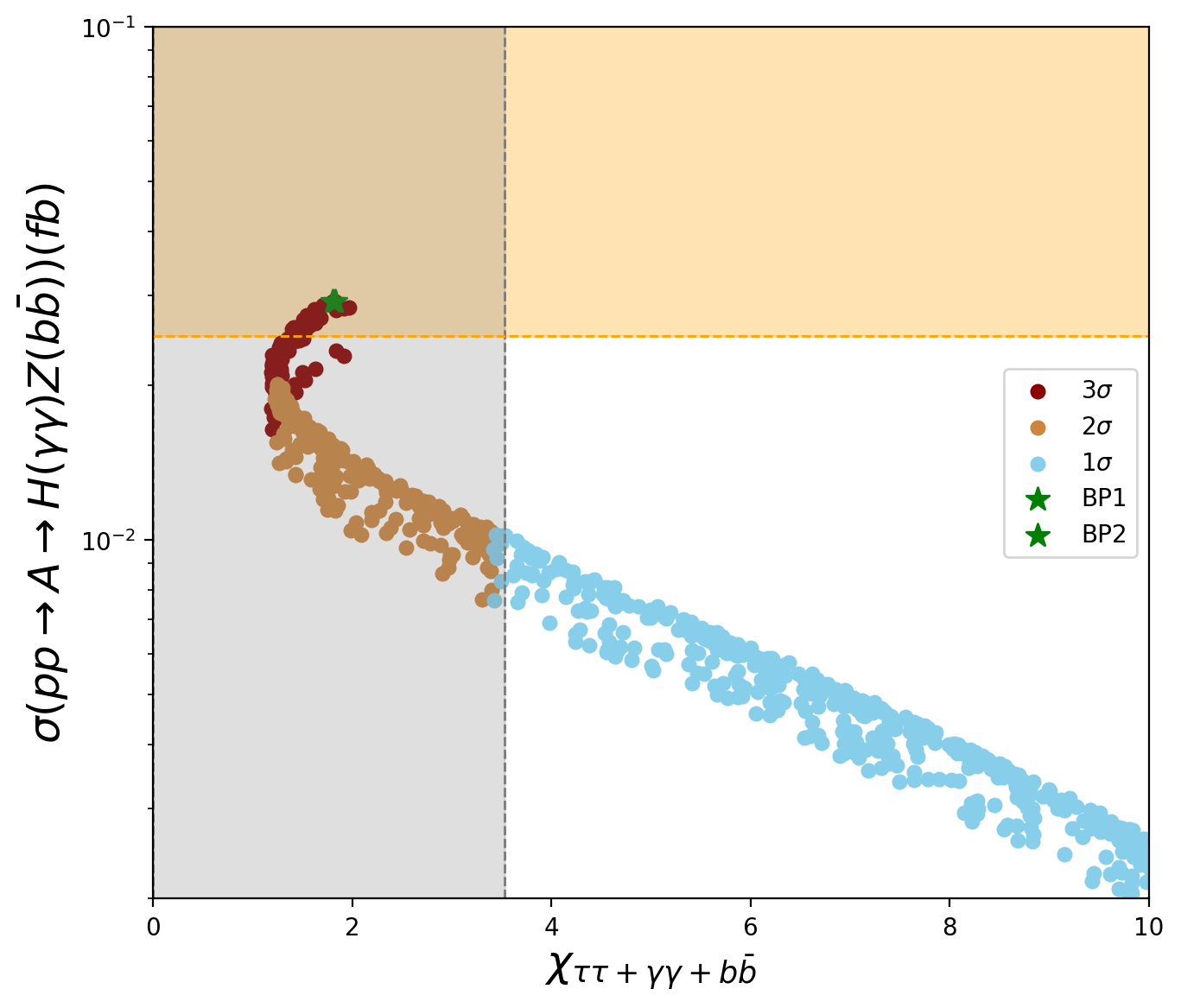}
    \caption{Scatter plot of $\chi_{\tau^+\tau^-+\gamma\gamma+b\bar{b}}$ versus the cross section for the channel $\sigma(pp\rightarrow A \rightarrow H(\to\gamma\gamma)Z(\to b\bar{b}))(fb)$ for parameter points satisfying all theoretical and experimental constraints. Points in blue, orange, and red correspond to those passing the HiggsSignal $\chi^2$ goodness-of-fit test within $1\sigma$, $2\sigma$ and $3\sigma$, respectively. The grey dashed band denotes the $1\sigma$ limit on $\chi^2$ while the orange dashed band represents the $2.5\sigma$ limit from the CMS measurement of the analysed process. The BPs summarised in Table \ref{tab:simul_bench} are marked with a star.}
    \label{fig:sigmavschi}
\end{figure}

\begin{figure}
    \centering
    \includegraphics[scale=0.33]{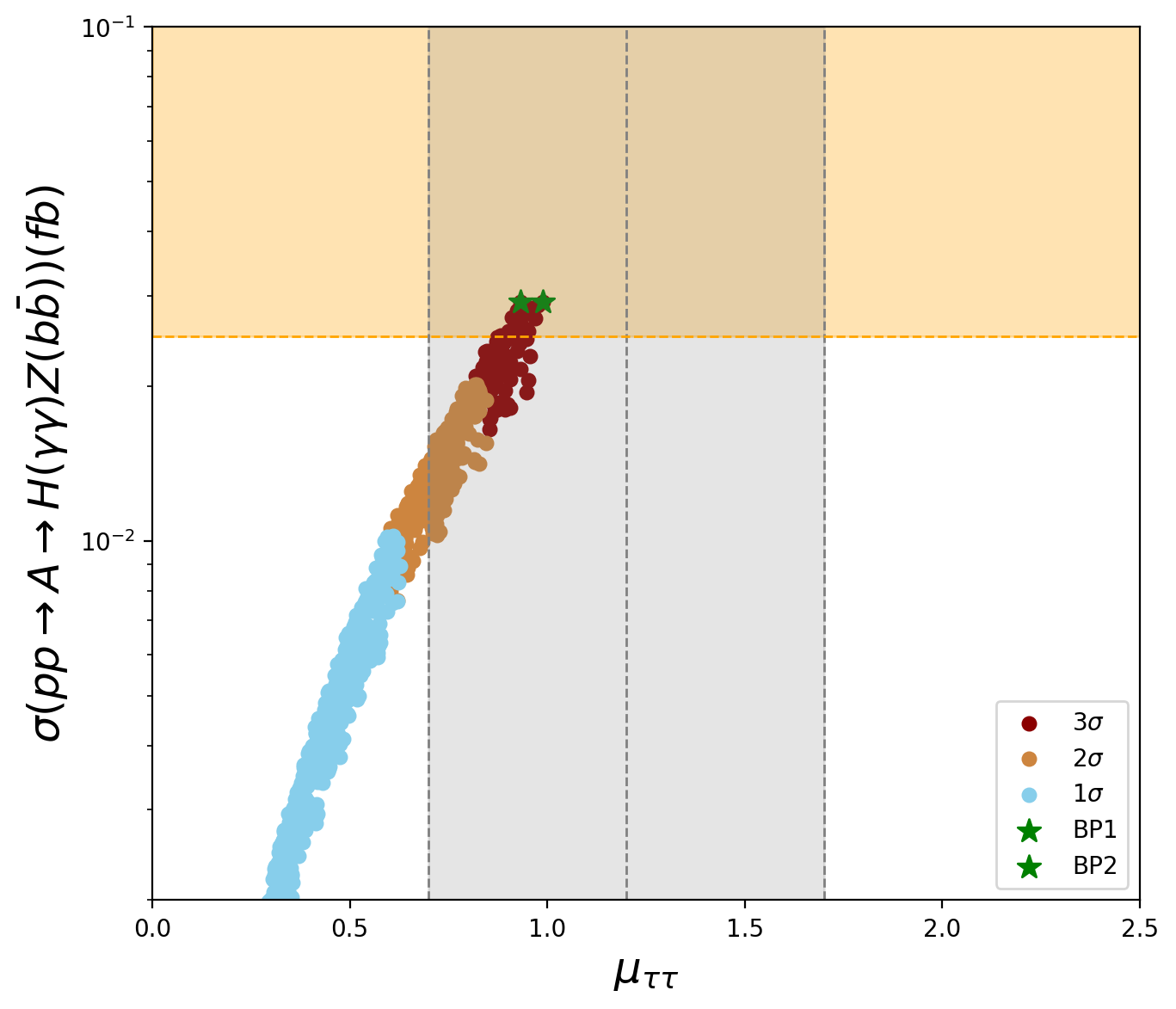}
    \includegraphics[scale=0.33]{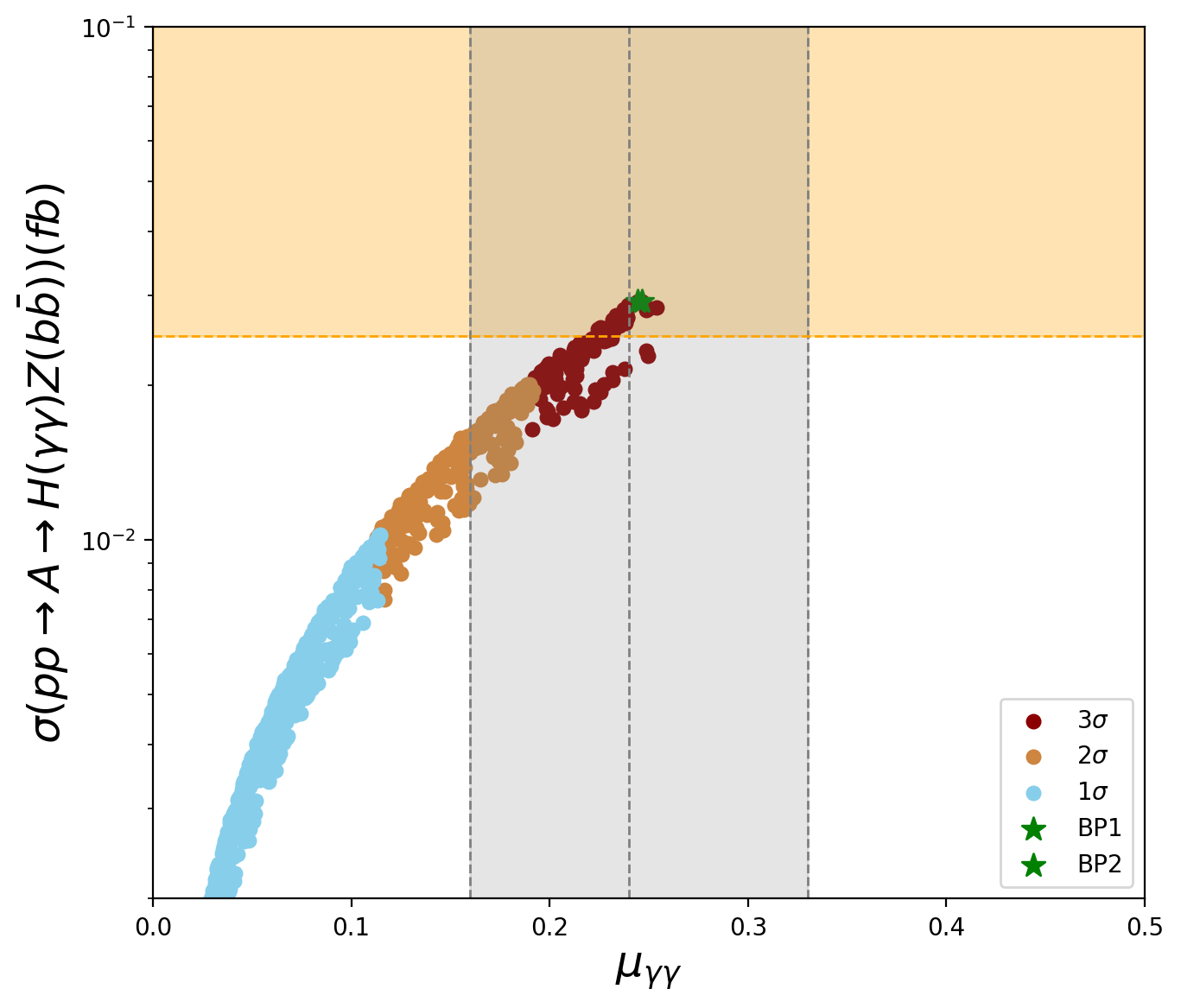}
    \includegraphics[scale=0.33]{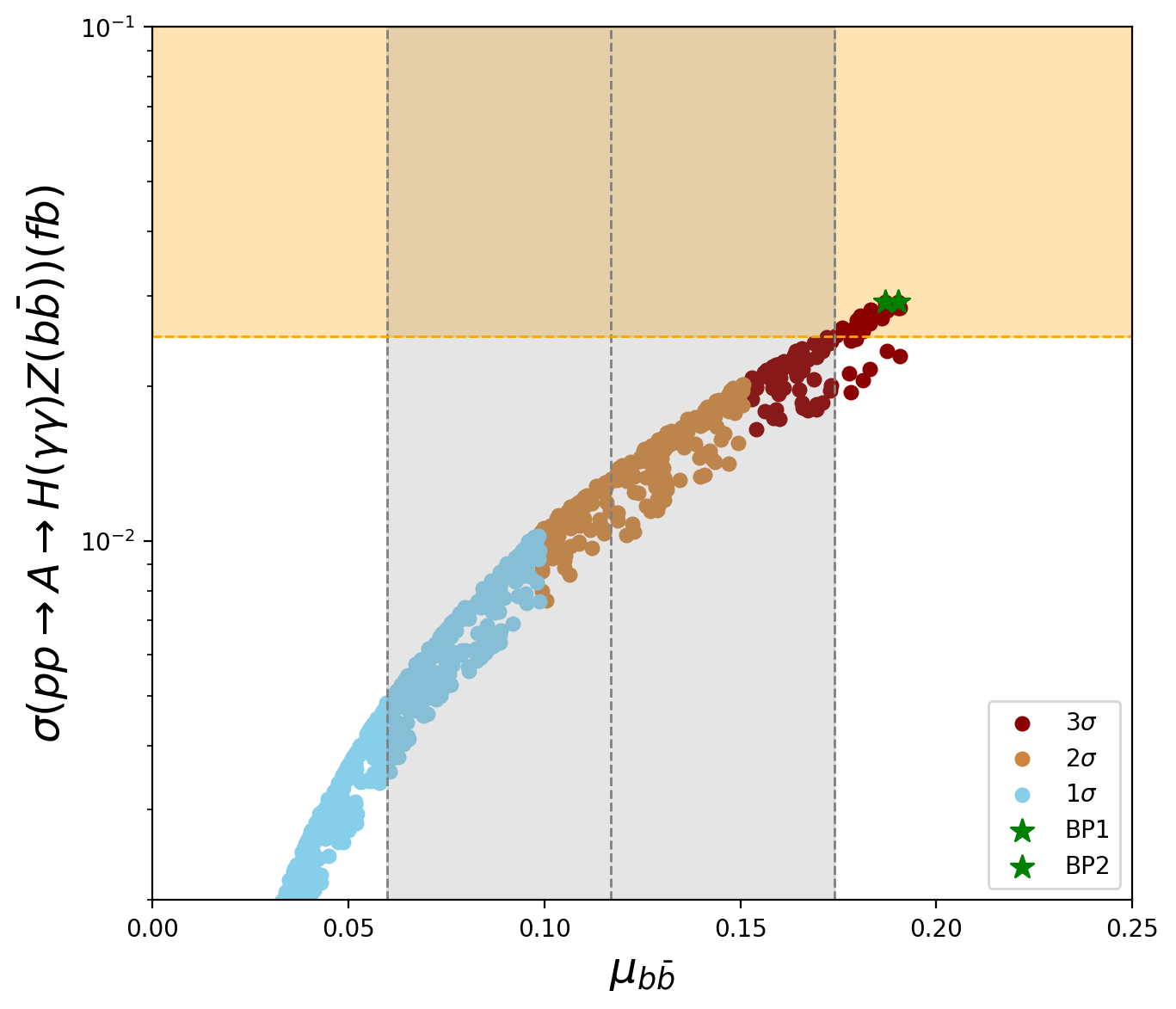}
    \caption{Scatter plot of the signal strengths in the three channels $\tau^+\tau^-$, $\gamma\gamma$ and $b\bar{b}$ versus the cross section. The colour scheme is consistent with that used in Figure \ref{fig:sigmavschi}. The BPs summarised in Table \ref{tab:simul_bench} are marked with a star.}
    \label{fig:sigmavschiseparate}
\end{figure}

Figure~\ref{fig:sigmavschi} displays the distribution of $\chi^2$ values as a function of the cross section $\sigma(pp\rightarrow A \rightarrow H(\to\gamma \gamma)Z(\to b \bar{b}))$. The points are colour-coded according to their agreement with the $\chi^2$ goodness-of-fit test: blue corresponds to $1\sigma$, orange to $2\sigma$ and red to $3\sigma$. The grey dashed band highlights the parameter space consistent with the $95$ GeV anomaly at the $1\sigma$ level while the orange dashed band marks the region consistent with the $650$ GeV anomaly at $2.5\sigma$.  
Our results therefore demonstrate that the 2HDM-I can successfully accommodate both excesses simultaneously, reproducing the $650$ GeV anomaly at $2.5\sigma$ and the $95$ GeV anomaly at $1\sigma$. The Benchmark Points (BPs) yielding the best simultaneous fits are summarised in Table~\ref{tab:simul_bench}. The signal strengths for the $\gamma\gamma$, $\tau^+\tau^-$ and $b\bar{b}$ channels are presented separately in Figure~\ref{fig:sigmavschiseparate}. This figure shows that a simultaneous agreement within $1\sigma$ can be achieved for the di-photon and di-tau channels, when considered independently, while such consistency is not realised for the $b\bar{b}$ channel. The colour scheme used in this plot is consistent with that of Figure~\ref{fig:sigmavschi}.

\begin{table}[h!]
    \centering
    \begin{tabular}{l l}
    \toprule[1pt]
        Parameter & \ \ \ \ \ \ \ Scan Range \\
        \midrule[1pt]
        $m_H$ & \ \ \ \ \ \ \ $125$ GeV \\
        $m_h$ & \ \ \ \ \ \ \  $93~\text{GeV}-97~\text{GeV}$ \\
        $m_A$ & \ \ \ \ \ \ \ $625~\text{GeV}-665~\text{GeV}$  \\
        $m_{H^\pm}$ & \ \ \ \ \ \ \ $620~\text{GeV}-675~\text{GeV}$  \\
        $\tan{\beta}$ & \ \ \ \ \ \ \ $2.3-10$  \\
        $\sin({\beta - \alpha})$ & \ \ \ \ \ \ \ $0 - 0.75$ \\
    \bottomrule[1pt]
    \bottomrule[1pt]
    \end{tabular}
    \caption{The scan ranges which are used for the MC sampling.}
    \label{tab:scanrange}
\end{table}

\begin{table}[h!]
    \raggedright
    \hspace*{-1.0cm}
        \hspace*{-0.75 truecm}
    \begin{tabular}{l l l l l l l l l l l l l l}
    \toprule[1pt]
        Parameter & \ \ \ \ \ \ \ $m_H$ & \ \ \ \ $m_h$ & \ \ \ \ $m_A$ & \ \ \ \ $m_{H^\pm}$  & \ \ \ \ $\tan \beta$ & \ \ \ \ $\sin(\beta - \alpha)$ & \ \ \ \ $\mu_{\tau^+ \tau^-}$ & \ \ \ \ $\mu_{\gamma \gamma}$ & \ \ \ \ $\mu_{b \bar{b}}$ & \ \ \ \ $\chi^2_{\tau^+ \tau^- + \gamma \gamma + b \bar{b}}$ & \ \ \ \ $\sigma_{pp \rightarrow A \rightarrow H(\gamma\gamma)Z(b \bar{b})}$ (fb)\\
        \midrule[1pt]  \\
         BP1 & \ \ \ \ 125.0 & \ \ \ \ 96.02 & \ \ \ \ 625.65 & \ \ \ \ 638.24 & \ \ \ \ 2.96 & \ \ \ \ 0.46 & \ \ \ \ 0.932 & \ \ \ \ 0.246 & \ \ \ \ 0.187 & \ \ \ \ 1.802 & \ \ \ \ 0.029\\
         BP2 & \ \ \ \ 125.0 & \ \ \ \ 93.30 & \ \ \ \ 627.88 & \ \ \ \ 633.04 & \ \ \ \ 2.95 & \ \ \ \ 0.46 & \ \ \ \ 0.988 & \ \ \ \ 0.244 & \ \ \ \ 0.190 & \ \ \ \ 1.830 & \ \ \ \ 0.029\\
    \bottomrule[1pt]
    \bottomrule[1pt]
    \end{tabular}
    \caption{BPs extracted. Masses corresponding to BSM scalars are in GeV unit.}
    \label{tab:simul_bench}
\end{table}

\section{Conclusions}
\label{sec:conclusions}

In this analysis, we have probed  the 2HDM-I as a possible theoretical framework addressing several data anomalies that have recently appeared in experimental data, most probably emerging from an extended Higgs sector, i.e., above and beyond the SM one. In particular, excesses over predictions within the latter have been seen in the search for new Higgs bosons with masses both below and above 125 GeV (the one of the SM-like Higgs state detected at the LHC in 2012). Excesses have been seen around 95 GeV in analyses targeting $e^+e^-\to Z\to  h(\to b\bar b) Z(\to {\rm jets})$ at LEP and $pp\to h\to \gamma\gamma$ and $\tau^+\tau^-$ at the LHC. In addition, a rather significant excess has been more recently detected also at 650 GeV, again, at the LHC, potentially due to $gg\to A\to  H(\to \gamma\gamma) Z(\to b\bar b)$. All these production and decay channels are present in our theoretical construct and, over large expanses of the parameter space of it, they can explain all aforementioned anomalies  at the 2.5$\sigma$ CL.

\section*{Acknowledgments}
The work of~S.M. is supported in part through the NExT Institute and  STFC CG ST/X000583/1. A.S. thanks the Anusandhan
National Research Foundation (ANRF) for providing financial support through the SERB-NPDF grant
(Ref No: PDF/2023/002572). A.K. acknowledges the support from Director's Fellowship at IIT Gandhinagar. 

\bibliography{bibliography}

\end{document}